# Ordering of Dopants and Potential Increase in $T_c$ to near Room Temperature

S.A.Wolf* and V.Z.Kresin**

*Departments of Physics and Materials Science, University of Virginia

Charlottesville, VA  22904

and

Office of the Assistant Secretary of Defense for Research and Engineering,

875 N Randolph St, Arlington VA 22203.

** Lawrence Berkeley National Laboratory, University of California at

Berkeley, CA 94720

Abstract

This paper describes a novel method to increase the resistive $T_c$ of cuprate superconductors to values approaching room temperature. The method is based on our analysis of the inhomogeneous superconducting state for the "pseudogap" region and on the interplay of doping and pair-breaking phenomena.  What we propose involves specific ordering of the dopants that leads to the formation of regions that can carry a supercurrent at high temperatures.

This paper describes a method to synthesize novel superconducting materials and structures with high values of the critical temperature. The high $T_c$ cuprates [1-3] continue to be a puzzling family of compounds.  One can show, and this is the topic of this Report that the structure of the cuprates could be modified and this modification would result in a noticeable increase in Tc.  Our analysis is based on the concept ,used to describe the pseudogap state , [ 4-7 ] developed by us jointly with Y.Ovchinnikov.

As we know, the region above the resistive Tc is characterized by peculiar properties (the pseudogap state). The most interesting region is the so-called underdoped regime. Since $T>T_c$ and the resistance is finite, one can imply that this state is normal, not superconducting. But, nevertheless, this state is characterized by a number of peculiar properties that are similar to those in superconductors. Two of them are especially important. The first is the presence of an energy gap, and the second which is even more essential, is the presence of an anomalous diamagnetism ("Meissner" effect) observed using different methods in [ 8,9 ] and recently in [10 ].

The observed coexistence of normal resistance and diamagnetism has been explained by us invoking the intrinsic inhomegeneity of the sample; we call it " intrinsic" because it is related to a fundamental process, namely to the doping (see below). The sample above $T_c$ is characterized by the presence of superconducting clusters (2D "islands") embedded into a normal metallic matrix. The matrix is responsible for the normal resistance, whereas the "islands"
provide the diamagnetic response. The presence such
an inhomogeneous structure was directly observed by STM magnetic spectroscopy in LaSrCuO [8] (see also the review [6]). Recently, key studies were also performed on $Bi_2Sr_2CaCu_2O_{8+\delta}$ samples in [11,12] and YBCO in [13]. All of these studies provide strong support for this picture (see also review [7]).

The superconducting regions appear below some characteristic temperature $T_c^*$ (we call it an "intrinsic"
critical temperature); its value greatly exceeds the usual resistive $T_c \equiv T_{c;res.}$. For example, for the underdoped LaSrCuO compound ($T_c \approx$ 25K) the value of $T_c^* \approx$80K. For the YBCO compound the value of $T_c^* \approx$250K, that is, it is close to room temperature. Decreasing the temperature towards $T_c \equiv T_{c;res}$ leads to an increase in the number of "islands" and to an increase in their size. The transition at $T_c$ is of percolative nature and corresponds to the formation of a macroscopic superconducting phase ("infinite cluster" in the terminology of the percolation theory) capable of carrying a macroscopic supercurrent. It is obvious that the superconducting clusters embedded in a normal matrix structure is unable to carry a superconducting current. Such a transfer of supercurrent might occur only at temperatures close to Tc. For example, the so-called "giant proximity effect" observed in [14] could be explained [5] by the formation of a Josephson tunneling network (SNSNS) which contains the superconducting islands.

It would be attractive to "order" the superconductive region and instead of statistically distributed superconducting "islands" to have a continuous superconducting phase; then one can expect the supercurrent to flow at high temperatures. In other words, it would mean an effective increase in the value of the resistive $T_c$. We are going to discuss the possibility of such an ordering, but initially one should stress one more important point, namely the origin of the observed inhomogeneity. As we know, the delocalized carriers, whose presence is responsible for the metallic and, correspondingly, the superconducting state, are created by doping. At the same time we should remember that the superconducting pairing could be depressed by the pair-breaking effect of impurities. The simplest and most well-known case of pair-breaking is the one provided by magnetic impurities [15], see also [16]. Furthermore, for the D-wave scenario pair-breaking can also be provided by usual, non-magnetic defects. For the cuprates the dopants represent such defects. Therefore, the dopants play a double role: they provide delocalized carriers and are also responsible for the pair-breaking. The statistical nature of doping leads to a random distribution of dopants and at relatively low doping the spatial distribution is rather broad. As a result, the regions with smaller number of dopants have larger value of the critical temperature; these regions form the superconducting "islands" inside of the normal matrix.

We think that an increase in $T_c$ can be achieved by a procedure that allows for a special ordering of defects. To clarify the concept, let us consider a specific example, namely, the YBCO compound. Doping of the parent $YBa_2Cu_3O_6$ sample is provided by adding oxygen into what would ultimately become the chain layer. The mixed-valence state of the in-plane Cu leads to the plane-chain charge transfer and the appearance of a hole, initially on the Cu site. Because of diffusion, the hole enters the system of delocalized carriers responsible for the metallic and, correspondingly, superconducting behavior: pairing of such holes causes superconductivity. The added oxygen ion and corresponding in-plane Cu form the defect mentioned above, with its pair-breaking impact.

We think that the picture could be strongly affected by specific ordering of the oxygen ions. More specifically, let us consider the thin film formed by a layer-by-layer deposition technique; the film is growing in the c-direction. The film should be build as a set of columns. The top layer could be imagined as a set of strips. Additional oxygen should be placed inside of specific columns. As a result, these columns would contain the chains with a composition close to $YBa_2Cu_3O_7$. The chains are parallel to the inter-strip boundaries. Such columns could be called the reservoir columns, since holes are created in these areas. The holes are delocalized and diffuse into the neighboring columns, which are free of defects and represent the high $T_c$ regions. Such a structure could be built, for example using a nano implantation technique (see,e.g., [17] with appropriate annealing or other methods for providing an ordered defect structure similar to what has been described here..

As a whole, the film would then contain alternating reservoir and high $T_c$ columns. Because of the presence of the oxygen defects, the reservoir columns have value of $T_c$ lower than in the neighboring columns that are free of defects. It is expected that the high $T_c$ columns would have values of $T_c$ close to the intrinsic critical temperature, $T_c^*$, that, in fact, could be close to room temperature. Moreover, such a structure would provide a continuous path for the supercurrent and this means an effective increase in the resistive $T_c$

The value of $T_c$ in these high $T_c$ columns could be depressed by the proximity effect with the reservoir columns, and this leads to some limitation on the width of the high $T_c$ columns, $W_H$. Indeed, we have the $S_L$-$S_H$ proximity system where L and H correspond to the reservoir and high $T_c$ columns ($T_{cH}$> $T_{cL}$). We know that the scale of the proximity effect is of order of the coherence length, $\xi_H$, of the high $T_c$ region. To minimize the impact of the proximity effect, the width of the high Tc column, $W_H$, should be larger than $\xi_H$. However the value of $\xi_H$ is rather small. Indeed, $\xi_H \approx \hbar v_F / 2\pi T_{c;H}$. If we take the values $V_F \approx 10^7$ sm/sec, and $T_c \approx (2\text{-}2.5)\,10^2$ K, we obtain $\xi_H \approx$ 5-10A. Because of such a small value of the coherence length, the condition $W_H >> \xi_H$ is perfectly realistic. Note that similar ordering could be effective for an increase in $T_c$ for other cuprates, e.g., LaSrCuO or Bi-2212 compounds.

The proposed method is conceptually analogous to the observed increase in $T_c$ for the cuprates caused by pressure (see, e.g., [18] and the analysis [19]), by an applied field [20], or by the photoinduced effect [21]. Indeed, the external pressure, radiation, or an applied field affects the doping without creating defects, that is, pair-breakers. The specifics of the structure proposed here is that the carriers appear in a region (high $T_c$ column) which is free of defects; they are produced in a different spatially separated column. As was stressed above, we observe the impact of such separation in the existing cuprates (in the "oseudogap state"), especially in the underdoped region, where the superconducting clusters (high $T_c$ regions) are embedded into a normal matrix. Thanks to such a separation, one can observe the anomalous diamagnetism (Meissner effect) at temperatures higher than the resistive $T_c$. The proposed ordering leads to a noticeable increase of the critical temperature for the resistive transition and thus a much higher effective transition temperature. An effect similar to what we have described here, has been observed in a $Sr_2CuO_{3+\delta}$ sample where magnetic transitions have been observed up to 95K when the oxygen dopants are ordered [22].

In summary, based on the concept [4-7] describing the pseudogap state, we propose a method of building a structure containing ordered defects. Thin films with such a structure containing regions (columns) with different compositions of dopants should be capable of carrying a supercurrent at temperatures greatly exceeding those for the existing cuprates.

The work of SAW is supported by The Office of the Assistant Secretary of Defense for Research and Engineering (ASDRE). The work of VZK is supported by AFOSR..